# The Photometric Periods of the Nova-Like Cataclysmic Variable LQ Pegasi (PG 2133+115)

Gerald D. Rude II and F. A. Ringwald*
Department of Physics
California State University, Fresno
2345 E. San Ramon Ave., M/S MH37
Fresno, CA 93740-8031

**ABSTRACT**

We present a time-resolved differential photometric study and time series analysis of the nova-like cataclysmic variable star LQ Peg. We discover three periodicities in the photometry, one with a period of 3.42 ± 0.03 hours, and another with a period of 56.8 ± 0.01 hours. We interpret these to be the apsidal superhump and precessional periods of the accretion disk, respectively, and predict that the orbital period of LQ Peg is 3.22 ± 0.03 hours. The third periodicity, with a period of 41.3 ± 0.01 hours, we interpret to be the nodal precessional period of the accretion disk. We also report a flare that lasted four minutes and had an energy in visible light of $(1.2 \pm 0.3) \times 10^{36}$ ergs, or $10^{4\text{-}5}$ times more energetic than the largest solar flares, comparable to the most energetic visible-light stellar flares known. We calculate the absolute magnitude of LQ Peg to be $M_J = 4.78 \pm 0.54$, and its distance to be 800 ± 200 pc.

\* Corresponding author. Tel: +1-559-278-8426; fax: +1-559-278-7741.
*E-mail address*: `ringwald@csufresno.edu`

## INTRODUCTION

Cataclysmic variables (hereafter CVs) are close binary star systems in which a late-type star, usually approximately on the main sequence, fills its Roche lobe and spills gas onto a white dwarf through an accretion disk, although the disk may be disrupted if the white dwarf has a strong magnetic field. Reviews of CVs include those of Robinson (1976), Warner (1995), and Heller (2001).

LQ Pegasi was discovered during the Palomar-Green Survey (Green et al., 1986), in which it was designated PG 2133+115 and classified as a CV. It was not listed in their preliminary report on CVs found by the survey, however (Green et al., 1982). Ferguson et al. (1984) presented visible and ultraviolet spectra that showed weak emission lines inside broad absorption wings on a blue continuum. They classified LQ Peg as a "thick-disk" or "UX UMa" cataclysmic variable, since the spectra they showed were similar to those of dwarf novae in outburst, or nova-like variables, which are thought to resemble dwarf novae stuck in outburst most of the time.

LQ Peg was listed as "Peg6" by Downes and Shara (1993), who gave precise coordinates and a finding chart: so do Downes et al. (1997) and Downes et al. (2001). It was given the variable star name LQ Pegasi by Kazarovets and Samus (1997).

Ringwald (1993) attempted to measure the orbital period of LQ Peg by doing a radial velocity study, but obtained only an uncertain estimate of 2.9 hours (0.121 ± 0.001 days). This was uncertain because the time series was short, which gave problems with aliasing (see Thorstensen and Freed, 1985). Another reason this orbital period is suspect is due to the difficulty in measuring the H$\alpha$ line. The H$\alpha$ line was quite narrow and quite faint, relative to the continuum, during the observations. This resulted in the only orbital-period measurement done by a radial-velocity study of a CV by Ringwald (1993) in which there was a significant false-alarm probability, of 17.4%, as defined by Thorstensen and Freed (1985). This means that there was a non-negligible chance that the velocity measurements themselves showed a false signal, regardless of aliasing, which was determined by how they were sampled.

Misselt and Shafter (1995) obtained six hours of time-resolved, *V*-band photometry of LQ Peg. The only variability they noted was the flickering that is characteristic of all CVs while transferring mass. (See Chapter 10 of Hellier, 2001.) A photometric study by Papadaki et al. (2006) concluded that LQ Peg has an orbital period of 2.99 hours (0.124747 ± 0.000006 days), although, as we will show, we suspect this is not the orbital period.

Ak et al. (2008) assumed that the orbital period of 0.12475 days was correct, and used it and *JHK$_s$* magnitudes observed for LQ Peg by the 2MASS survey (Cutri et al., 2003) to estimate an absolute magnitude of $M_J$ = 4.76, a distance of 809 pc, and *E(B-V)* = 0.095, estimated from the position in the Galaxy and the distance that Ak et al. (2008) found for LQ Peg. Even if this assumed orbital period is incorrect and these values are only approximate, they are still consistent with LQ Peg being a relatively luminous, nova-like CV.

LQ Peg goes into unpredictable low states, similar to those of VY Sculptoris and similar stars, which are sometimes called "anti-dwarf novae" (Kato and Uemura, 1999; Schmidtke et al., 2002; Honeycutt and Kafka, 2004; Kafka and Honeycutt, 2005; Shugarov et al., 2007). This further supports that LQ Peg is a nova-like CV. Honeycutt and Kafka (2004) give a light curve from 1993 to 2003. It shows that LQ Peg spends most of its time in the high state, as VY Scl stars do. For LQ Peg, this high state is near *V* = 14.6, but LQ Peg can drop to *V* = 17.0 during a low state.

**OBSERVATIONS**

Time-resolved CCD photometry of LQ Peg was collected on the nights of 2010 August 20-28 UT, with the exception of August 22 UT. We used the 0.41-m (16-inch) f/8 telescope by DFM Engineering at Fresno State's station at Sierra Remote Observatories and a Santa Barbara Instruments Group STL-11000M CCD camera. Frames were exposed for 23 seconds, with a dead time to read out the CCD between exposures of 7 seconds, making for a total time resolution of 30 seconds. All exposures were taken through a Clear luminance filter by Astrodon. Weather was clear and apparently photometric on all nights, although the third night (August 23 UT) may have had some haze. Table 1 is a journal of the observations.

Table 1: Journal of observations

| UT Date | UT Start | Duration (hr) | Number of exposures (23+7 s for all) |
|---|---|---|---|
| 2010 August 20 | 4:26 | 7.61 | 874 |
| 2010 August 21 | 3:40 | 8.3 | 947 |
| 2010 August 23 | 3:44 | 8.04 | 920 |
| 2010 August 24 | 3:25 | 8.36 | 962 |
| 2010 August 25 | 9:33 | 2.12 | 244 |
| 2010 August 26 | 8:54 | 2.34 | 270 |
| 2010 August 27 | 3:27 | 8.12 | 937 |
| 2010 August 28 | 3:37 | 7.88 | 897 |

The data were processed with AIP4WIN 2.0 software (Berry and Burnell, 2005). All exposures were dark-subtracted, but not divided by a flat field. The CCD temperature was −5° C for all exposures. Fifteen dark frames were collected every night. These frames were median combined to form a master dark frame, which was then subtracted from each of the target frames.

To measure the photometry, we used the same check stars as those used by Papadaki et al. (2006). Check stars S1 – S8 were used for ensemble photometry (Gilliland and Brown, 1988), with S4 being relabeled as C1 and S7 as C2. This telescope and camera have an image scale of 0.51 arcseconds/pixel. All our observations were done binned 3x3, making for an image scale of 1.53 arcseconds/pixel. The seeing ranged between 0.95 and 1.72 arcseconds on all nights. All photometry used an aperture of 6 pixels, or 3.06 arcseconds, in diameter, and used an annulus around between 9 and 12 pixels (4.59 and 6.12 arcseconds) in diameter around this, for sky subtraction.

Figure 1 shows a typical nightly light curve of our differential photometry of LQ Peg, comparison (C1), and check (C2) stars, from the first night (2010 August 20 UT). Figure 7 shows a light curve from the sixth night (2010 August 26 UT), since it showed an atypical flare. Light curves from all other nights are included in Figures 8-12 in the Appendix (available electronically only).

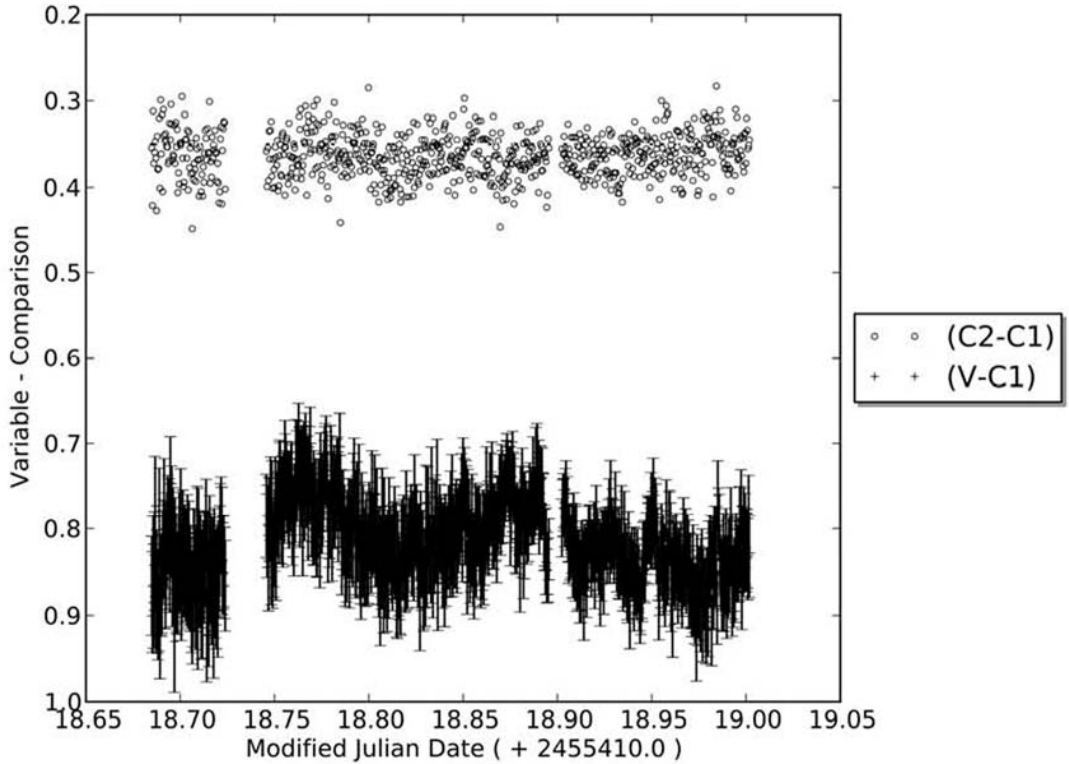

Figure 1: Differential Light Curve of LQ Peg for 2010 August 20 UT.

**ANALYSIS**

Figure 2 is a Lomb-Scargle periodogram (Lomb, 1976; Scargle, 1982; Press et al., 1992) of our time-resolved photometry that we calculated with the PERANSO (PeRiod ANalysis SOftware) software package (Vanmunster, 2009) to search for periodicities. Before calculating this periodogram, we cut out eight data points from the time series observed on August 26 UT, during which we suspect a flare erupted (see the section on this flare, below).

Figure 2 shows that the low frequency signal at 0.4162 cycles/day is by far the most prominent signal. Other significant periods were observed at: 0.5809, 1.4252, 2.0180, 3.0150, 7.0390, and 8.0240 cycles/day.

We note four distinct periodicities, accompanied by aliases. The first periodicity has a frequency $f_1$ = 0.4162 cycles/day, corresponding to a period $P_1$ = 56.8 ± 0.01 hours. The periodogram shows its one-cycle-per-day alias at a frequency of 1.42 cycles/day. The second periodicity has $f_2$ = 7.039 cycles/day, corresponding to a period $P_2$ = 3.42 ± 0.03 hours. The periodogram shows its one-cycle-per-day aliases at frequencies of 6.06 and 8.02 cycles/day. The third periodicity has $f_3$ = 0.5809 cycles/day, corresponding to a period $P_2$ = 41.3 ± 0.01 hours. The fourth periodicity has $f_4$ = 3.0 cycles/day, with aliases at 2.0 and 4.0 cycles/day.

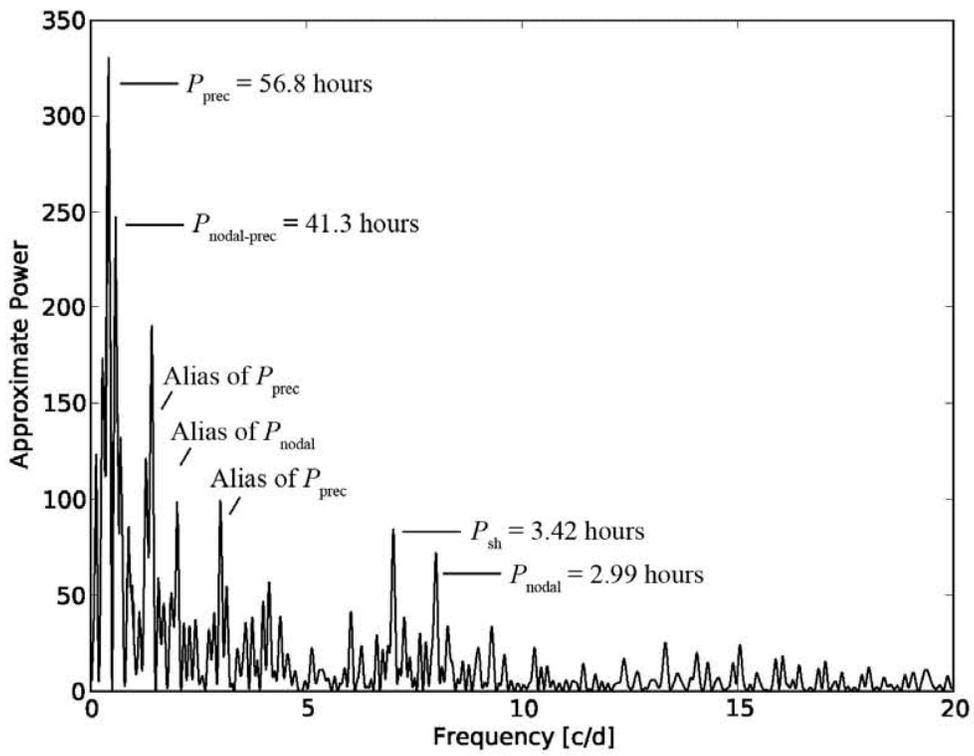

Figure 2: Lomb-Scargle periodogram showing the periods of note.

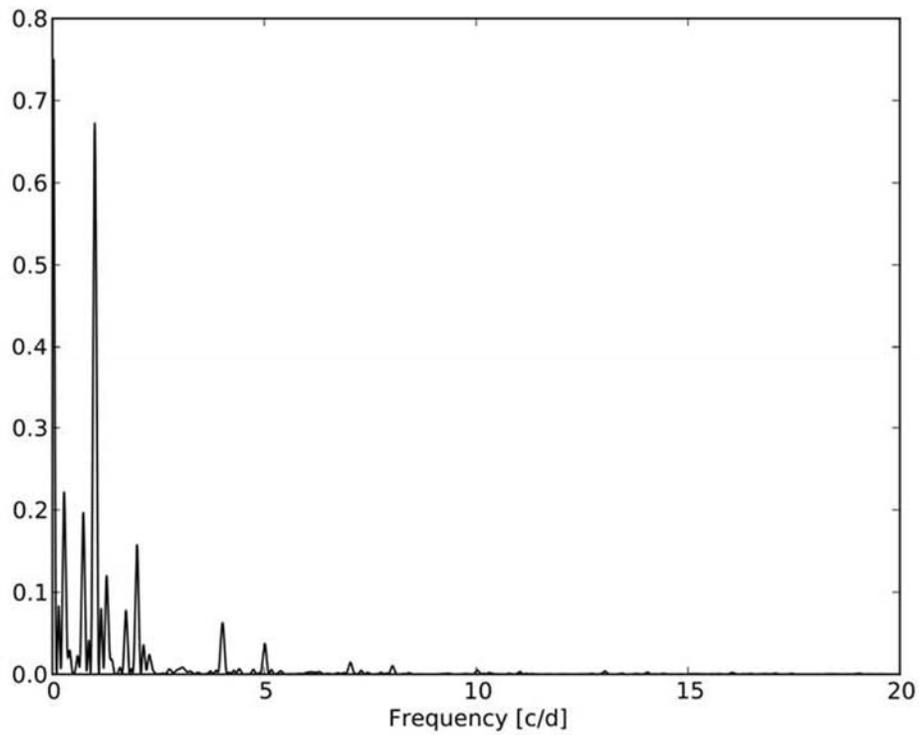

Figure 3: Spectral Window Function of LQ Peg.

Figure 3 shows the spectral window function for our observations. This plots the pattern caused by aliasing because of the convolution with the diurnal cycle (see the discussions of the convolution theorem and window functions by Press et al., 1992), and shows what signals could inadvertently be introduced to the data set because of the spacing of our observations. Signals found in our time series analysis, which coincide with signal peaks in the spectral window function, are regarded as suspect and ignored. We will therefore ignore the signal with $f_4$ = 3.0 cycle/day signal, as well as its aliases at 2.0 and 4.0 cycles/day.

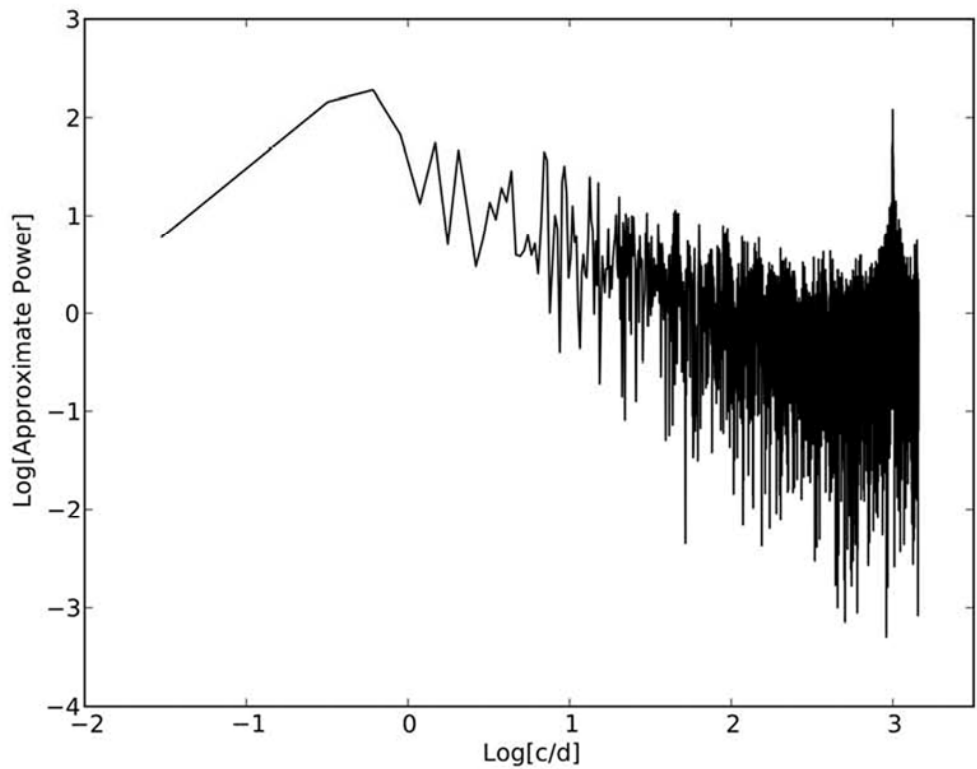

Figure 4: A log-log plot of the Lomb-Scargle periodogram shown in Figure 2. The signal at the tail end of this plot is an artifact of our sampling rate at the Nyquist frequency $f$, where log ($f$ cycles/day) = 3.16.

Figure 4 is a log-log plot of Figure 2, the Lomb-Scargle periodogram. It shows that the maximum power is at frequencies less than one cycle/day. It also shows no obvious excess of power at high frequencies, as can be caused by Quasi-Periodic Oscillations (QPOs), or Dwarf Nova Oscillations (DNOs). (See Chapter 10 of Hellier, 2001.) There is a power spike at the Nyquist Frequency which may be the result of our sampling rate or unresolved higher frequency periodicity.

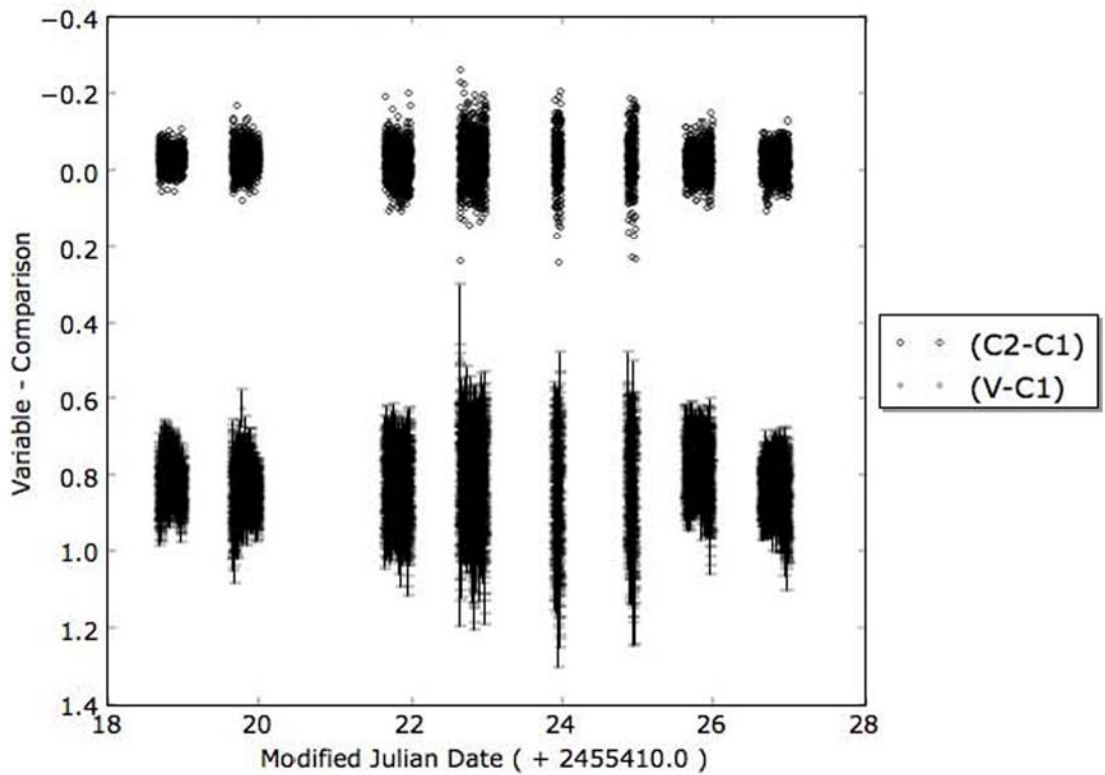

Figure 5: Differential Light Curve of LQ Peg

Figure 5 shows a fluctuation in the mean magnitude of LQ Peg. Phase folding the data over the 0.4162 cycles/day frequency (56.8-hour period), as shown by Figure 6, shows sinusoidal behavior.

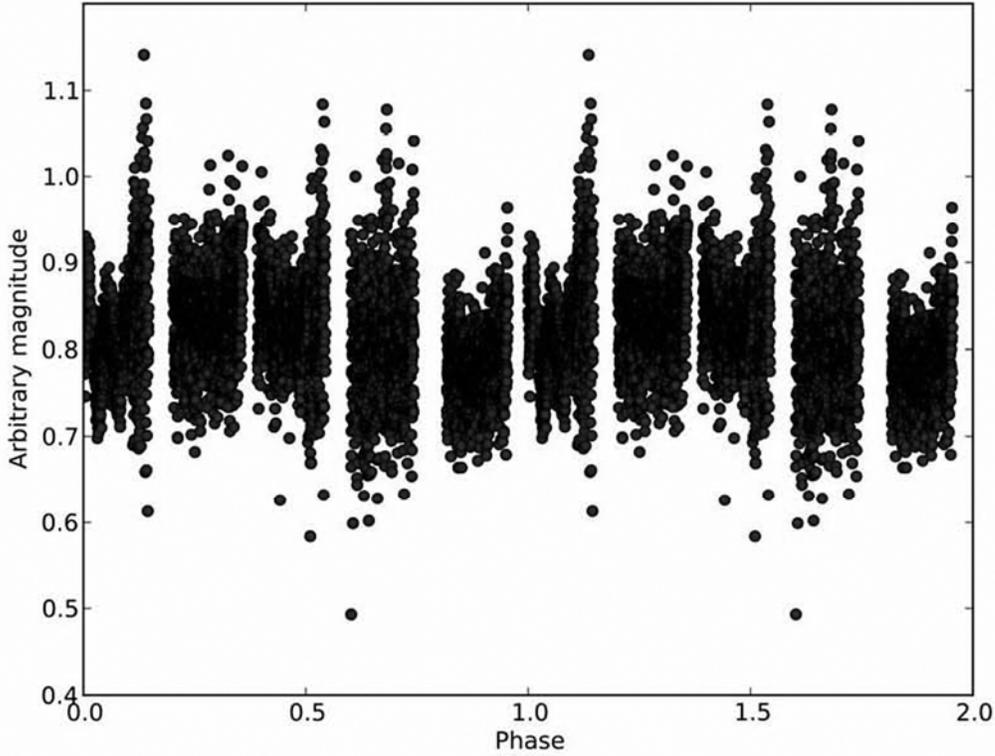

Figure 6: The data have been phase folded over the low-frequency (56.8-hour) signal.

Papadaki et al. (2006) normalized, or "pre-whitened" their time-resolved photometry, to eliminate night-to-night variations in the system's mean magnitude. Unsurprisingly, their time series analysis did not show the low-frequency signals we found.

Superhumps, both apsidal and nodal, can give rise to the periodicities observed in LQ Peg. As discussed in Chapter 6 of Hellier (2001), positive superhumps (also known as apsidal superhumps) are fluctuations in the light curve due to eccentricity in the accretion disk. When the precessional period of the elliptical disk and the orbital period of the star "beat" against one another, this periodic phenomenon can show in the light curve.

Interpreting the low-frequency signal of 0.4162 cycles/day ($P_1 = 56.8 \pm 0.01$ hours) to be associated with the precessional period $P_{prec}$, and the period associated with the signal at 7.039 cycles/day ($P_2 = 3.42 \pm 0.03$ hours) to be the apsidal superhump period $P_{sh}$, we predict that the orbital period $P_{orb} = 3.22 \pm 0.03$ hours, since (see page 77 of Hellier, 2001):

$$\frac{1}{P_{sh}} = \frac{1}{P_{orb}} - \frac{1}{P_{prec}}$$

Figure 6.19 of Hellier (2001) plots apsidal superhump period excess ($\varepsilon_{sh}$) versus orbital period in hours. Using our values for $P_{orb}$ and $P_{sh}$, we find a period excess $\varepsilon_{sh} = (P_{sh} - P_{orb})/P_{orb} = 0.056$, which fits the upper trend line of Figure 6.19 of Hellier (2001).

The orbital period is not obvious in our data, but LQ Peg may be nearly face-on, which may explain the difficulty in obtaining a radial velocity curve. A thick disk (Ferguson et al., 1984), characteristic of a luminous nova-like CV, is expected to be bright everywhere, not just where the gas stream from the mass-losing secondary star strikes the edge of the disk.

Taking the 0.5809 cycle/day signal (corresponding to $P_3 = 41.3 \pm 0.01$ hours) to be the nodal precessional frequency ($1/P_{nodal}$), and the orbital period to be $3.22 \pm 0.03$ hours, we arrive at a nodal superhump period $P_{ns} = 2.99 \pm 0.03$ hours. Harvey et al. (1995) called these "negative superhumps"; Hellier (2001) called them "infrahumps," and wrote their period as $P_{ih}$. They are thought to be from bending waves in the accretion disk, perpendicular to the disk's plane. This matches the signal observed by Papadaki et al. (2006), though they interpreted this as the orbital period. In our interpretation,

$$\frac{1}{P_{ns}} = \frac{1}{P_{orb}} + \frac{1}{P_{nodal}}$$

$$\varepsilon_{nodal} = -\frac{1}{2}\varepsilon_{sh} = -0.031$$

The nodal superhump excess $\varepsilon_{nodal}$ fits the lower trend line in Figure 6.19 of Hellier, 2001. This interpretation may explain why there is so much power in the peak at 8.02 cycles/day: this frequency is a one-cycle-per-day alias of the 3.22-hour apsidal superhump period, and it is also the nodal superhump period.

**REVISED ABSOLUTE MAGNITUDE AND DISTANCE**

The NASA/IPAC Extragalactic Database (NED) shows that $E(B-V) = 0.097$ in the direction of LQ Peg, at the distance estimated by Ak et al. (2008). The 2MASS survey (Cutri et al., 2003) gives $J = 14.383 \pm 0.040$, $H = 14.449 \pm 0.072$, and $K_s = 14.333 \pm 0.072$ for LQ Peg. Using the method of Ak et al. (2007), and assuming our predicted orbital period of 3.22 hours, we calculate the absolute magnitude $M_J = 4.78 \pm 0.54$ and distance of $800 \pm 200$ pc for LQ Peg. Assuming $V = 14.6$, this implies $M_V = 4.98 \pm 0.54$. These values are consistent with LQ Peg being a nova-like CV: the relation of Warner (1987) shows that, for a dwarf nova in outburst with the same orbital period, $M_V = 4.81$.

## THE FLARE OF 2010 AUGUST 26

A flare occurred on August 26 UT. It is shown in Figure 7. It was unlikely to have been from a cosmic ray hitting the CCD detector, since it was present in eight frames, over a total duration of four minutes. Following the method of Chugainov (1972), we calculated the equivalent duration of the flare $P = \Sigma \, [(I_f - I_0)/I_0] \, \Delta t$, where $I_f$ = total intensity of the star in its flare state minus sky background, $I_0$ = mean intensity of the star in quiescence minus sky background, and $\Delta t$ = the time resolution of the observations = 30 seconds. We found $P = 4.96$ minutes. We then calculated the total flare energy $E = P \, L_q$, where $L_q$ is the quiescent star luminosity band in visible light, using the absolute magnitude $M_v = 4.98 \pm 0.54$, estimated in the previous section.

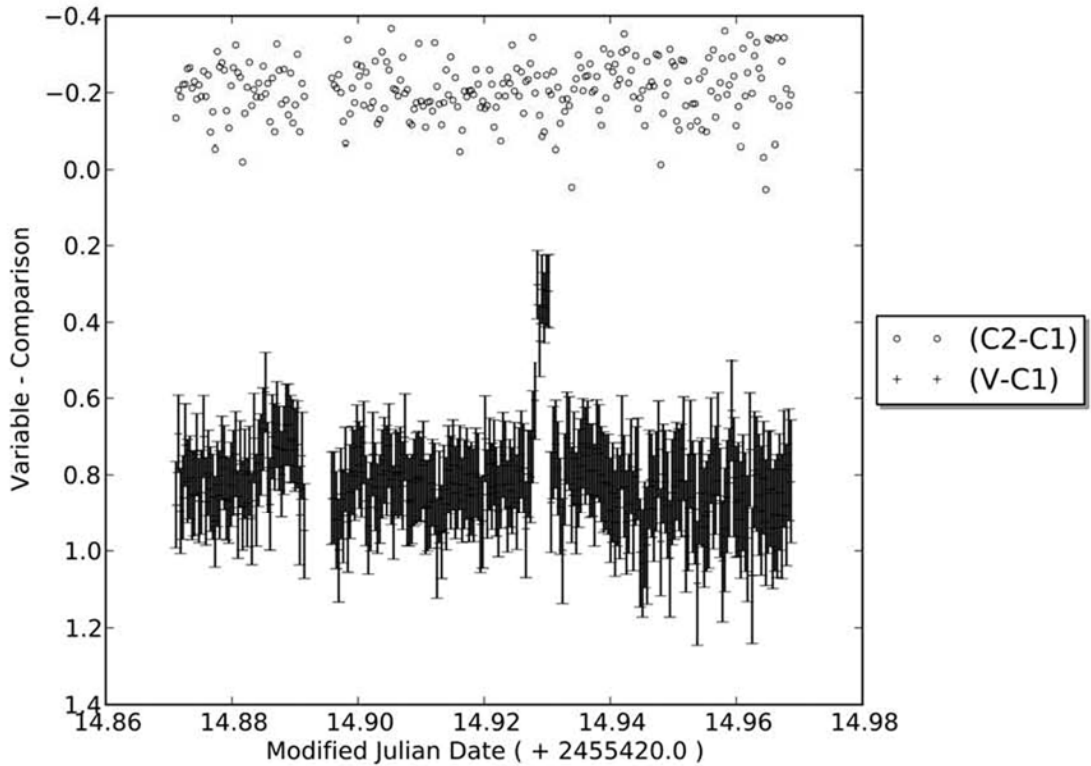

Figure 7: Differential Light Curve of LQ Peg for 2010 August 26 UT.

We estimate the flare's energy in visible light to have been $E = (1.2 \pm 0.3) \times 10^{36}$ ergs. This may be an overestimate, but probably not by more than a factor of two, since we observed this flare through a Clear filter that is transparent over a range of wavelengths about twice that of a *V* filter.

Even so, this flare was far more energetic than solar flares, which typically have energies of $\sim 10^{29}$ ergs in visible light, the largest of which have energies of $3 \times 10^{31}$ ergs in visible light and $2.5 \times 10^{31}$ ergs in X-rays. This flare was comparable in energy to flares in RS CVn binary star systems, which are the most energetic magnetic stellar flares known, with bolometric flare energies of up to $10^{38}$ ergs (Haisch et al., 1991), and flares in the *U* and *B* bands with energies of up to $1.8 \times 10^{35}$ ergs (Mathioudakis et al., 1992).

Schaefer et al. (2000) and Rubenstein and Schaefer (2000) pointed out a class of stars that may have flares this energetic, although RS CVn stars are usually more active in X-rays, the ultraviolet, and the radio, not in visible light (Jeffries and Bedford, 1990; Henry and Newsom, 1996). Also, RS CVn flares typically last hours or days, not minutes.

If this event was a magnetic flare, it may have been either on the secondary star or on the accretion disk. There is no indication that the white dwarf in LQ Peg is magnetic, since LQ Peg does not have strong He II 468.6 nm emission (Ferguson et al., 1984). NASA's HEASARC archive does not even list LQ Peg as an X-ray source, whereas it does for all the magnetic CVs and many non-magnetic CVs found by the Palomar-Green survey (Silber, 1992). We therefore consider it unlikely that this flare occurred because of a magnetic interaction between this CVs' two stars.

Low states in CVs, such as those observed in LQ Peg and other VY Scl stars, may be caused by magnetism on the CV's secondary star (Livio and Pringle, 1994). Because of all these considerations, we think it likely that the flare we observed occurred on the secondary star.

## CONCLUSIONS

We carried out a photometric study of LQ Pegasi. Time series analysis was performed and a low-frequency signal of period 56.8 hours, not discussed by the previous literature on the star, was discovered. Our analysis indicates that the low frequency periodicity appears to be the precessional period of both an elliptical accretion disk, and nodal precession of the disk. Based on this analysis, we predict that the true orbital period of LQ Peg is 3.22 hours, that the 3.42-hour period we have observed and reported in this paper is the apsidal superhump period, and that the 2.99 hour periodicity observed by Papadaki et al. (2006) is the nodal superhump period of the accretion disk.

To confirm our prediction, it will be necessary to carry out an improved radial velocity study of LQ Peg. We recommend tht the spectra of LQ Peg of Papadaki et al. (2006) and of Rodríguez-Gil et al. (2007) be re-phased over the 3.22-hour orbital period we have predicted. If this orbital period is correct, the time-resolved behavior and Doppler tomogram of LQ Peg may become clearer.

## ACKNOWLEDGEMENTS

This research used photometry taken at Fresno State's station at Sierra Remote Observatories. We thank Dr. Greg Morgan, Dr. Melvin Helm, Dr. Keith Quattrocchi, and the other SRO observers for creating this fine facility, and we thank the Department of Physics at California State University, Fresno for supporting it.

This research has made use of the Simbad database, which is maintained by the Centre de Données astronomiques de Strasbourg, France. This research has made use of NASA's Astrophysics Data System. This publication makes use of data products from the Two Micron All Sky Survey, which was a joint project of the University of Massachusetts and the Infrared Processing and Analysis Center/California Institute of Technology, funded by the National Aeronautics and Space Administration and the National Science Foundation. This research has made use of the NASA/IPAC Extragalactic Database (NED), which is operated by the Jet Propulsion Laboratory, California Institute of Technology, under contract with the National Aeronautics and Space Administration. This research has made use of data and software provided by the High Energy Astrophysics Science Archive Research Center (HEASARC), which is a service of the Astrophysics Science Division at NASA/Goddard Space Flight Center and the High Energy Astrophysics Division of the Smithsonian Astrophysical Observatory.